\newcommand{\p}{\partial}
\newcommand{\f}[2]{\frac{#1}{#2}}

\newcommand{\pder}[2]{\frac{ \p #1}{\p #2}}
\documentclass[preprint]{mn2e}
\def\jnl@style{\it}
\def\aaref@jnl#1{{\jnl@style#1}}

\def\aaref@jnl#1{{\jnl@style#1}}

\def\aj{\aaref@jnl{AJ}}                   
\def\araa{\aaref@jnl{ARA\&A}}             
\def\apj{\aaref@jnl{ApJ}}                 
\def\apjl{\aaref@jnl{ApJ}}                
\def\apjs{\aaref@jnl{ApJS}}               
\def\ao{\aaref@jnl{Appl.~Opt.}}           
\def\apss{\aaref@jnl{Ap\&SS}}             
\def\aap{\aaref@jnl{A\&A}}                
\def\aapr{\aaref@jnl{A\&A~Rev.}}          
\def\aaps{\aaref@jnl{A\&AS}}              
\def\azh{\aaref@jnl{AZh}}                 
\def\baas{\aaref@jnl{BAAS}}               
\def\jrasc{\aaref@jnl{JRASC}}             
\def\memras{\aaref@jnl{MmRAS}}            
\def\mnras{\aaref@jnl{MNRAS}}             
\def\pra{\aaref@jnl{Phys.~Rev.~A}}        
\def\prb{\aaref@jnl{Phys.~Rev.~B}}        
\def\prc{\aaref@jnl{Phys.~Rev.~C}}        
\def\prd{\aaref@jnl{Phys.~Rev.~D}}        
\def\pre{\aaref@jnl{Phys.~Rev.~E}}        
\def\prl{\aaref@jnl{Phys.~Rev.~Lett.}}    
\def\pasp{\aaref@jnl{PASP}}               
\def\pasj{\aaref@jnl{PASJ}}               
\def\qjras{\aaref@jnl{QJRAS}}             
\def\skytel{\aaref@jnl{S\&T}}             
\def\solphys{\aaref@jnl{Sol.~Phys.}}      
\def\sovast{\aaref@jnl{Soviet~Ast.}}      
\def\ssr{\aaref@jnl{Space~Sci.~Rev.}}     
\def\zap{\aaref@jnl{ZAp}}                 
\def\nat{\aaref@jnl{Nature}}              
\def\iaucirc{\aaref@jnl{IAU~Circ.}}       
\def\aplett{\aaref@jnl{Astrophys.~Lett.}} 
\def\apspr{\aaref@jnl{Astrophys.~Space~Phys.~Res.}}
\def\bain{\aaref@jnl{Bull.~Astron.~Inst.~Netherlands}} 
\def\fcp{\aaref@jnl{Fund.~Cosmic~Phys.}}  
\def\gca{\aaref@jnl{Geochim.~Cosmochim.~Acta}}   
\def\grl{\aaref@jnl{Geophys.~Res.~Lett.}} 
\def\jcp{\aaref@jnl{J.~Chem.~Phys.}}      
\def\jgr{\aaref@jnl{J.~Geophys.~Res.}}    
\def\jqsrt{\aaref@jnl{J.~Quant.~Spec.~Radiat.~Transf.}}
\def\memsai{\aaref@jnl{Mem.~Soc.~Astron.~Italiana}}
\def\nphysa{\aaref@jnl{Nucl.~Phys.~A}}   
\def\physrep{\aaref@jnl{Phys.~Rep.}}   
\def\physscr{\aaref@jnl{Phys.~Scr}}   
\def\planss{\aaref@jnl{Planet.~Space~Sci.}}   
\def\procspie{\aaref@jnl{Proc.~SPIE}}   

\usepackage{epsfig}
\begin{document}
\title{The role of General Relativity in the
  evolution of Low Mass X-ray Binaries}

\author[G. Lavagetto et al.]{G. Lavagetto$^1$\thanks{email: lavaget@fisica.unipa.it}, L. Burderi$^2$, 
F. D'Antona$^2$,
T. Di Salvo$^1$, 
R. Iaria$^1$,
and N. R. Robba$^1$
\\
$^1$ Dipartimento di Scienze Fisiche ed Astronomiche, 
Universit\`a di Palermo, via Archirafi n.36, 90123 Palermo, Italy.\\
$^2$ Osservatorio Astronomico di Roma, Via Frascati 33, 
00040 Monteporzio Catone (Roma), Italy.
}
\maketitle
\begin{abstract}
We study the evolution of Low Mass X-ray Binaries hosting a neutron
star and of millisecond binary radio
pulsars, with   the help of numerical simulations
that keep into account the detailed evolution of the companion star,
of the binary  system and of the neutron star. According to general relativity, when energy is released  during
accretion or due to magnetodipole radiation during the pulsar phase,
the system loses gravitational mass. Moreover, the neutron star can
collapse to a black hole if its mass exceeds a critical limit, that
depends on the equation of state of ultradense matter and is
typically $\sim 2 M_\odot$. These facts have some interesting
consequences:
1) In a millisecond radio pulsar the mass-energy is lost with a specific angular momentum that is
smaller than the specific angular momentum of the system, resulting in
a positive contribution to the orbital period derivative. 
If this contribution is dominant and can be measured, we can extract
information about 
the moment of inertia of the neutron star, since the energy loss rate depends
on it. Such a measurement can therefore help to put constraints  on
the equation of state of ultradense matter. 
2)In Low Mass X-ray Binaries  below the bifurcation period ($\sim 18$ h), the
neutron star survives the ``period gap'' only if its mass is smaller
than the maximum non-rotating mass when the companion becomes fully
convective  and accretion
pauses. Since
in  such evolutions $\sim 0.8 M_\odot$ can be accreted onto the
neutron star, short period ($P\le 2 h$) millisecond X-ray pulsar like
SAX~J1808.4-3658 can be formed only if either a large part of the
accreting matter has been ejected from the system, or the equation of
state of ultradense matter is very stiff.
3) In Low Mass X-ray binaries above the bifurcation period, the
mass-energy  loss lowers the
mass transfer rate. As side effect, the inner core of the companion
star becomes $\sim 1\%$ bigger than in a system with a non-collapsed
primary. Due to this difference, the final orbital period of the
system becomes $20\%$ larger than what is obtained if the mass-energy
loss effect is not taken into account.

\end{abstract}
\begin{keywords}
Stars: neutron -- X-rays: binaries -- binaries: close --
pulsars:general -- relativity -- stars: individual: SAX J1808.4-3658
\end{keywords}

\section{Introduction}

Low-Mass X-Ray Binaries (LMXBs) are systems consisting of a neutron
star (NS) with a relatively weak magnetic field ($< 10^{10}$ G) 
accreting from a low mass ($\sim 1 M_\odot$) companion star.
 When the companion star fills its Roche lobe, it transfers mass to
 the NS.
The companion fills its Roche lobe  either because it expands due to nuclear  evolution 
or because the lobe shrinks due to orbital angular momentum losses caused by gravitational
radiation  and
magnetic braking.
 The matter flowing from
the inner  lagrangian point towards the NS 
forms a Keplerian  accretion disc around it. The NS is spun
up by the  accreting matter, 
to an equilibrium period that is roughly equal to the keplerian
frequency at the  inner rim of the accretion disc \cite{gola79}.
Once accretion ends, the NS can light up as a fastly rotating
magnetodipole (radio pulsar): this is the so-called recycling scenario
for the formation of millisecond pulsars \cite{bhva91}.

The secular evolution of LMXBs can follow two very different paths
according to the evolutionary stage of the companion at the start of
the mass transfer.  If the orbital
period of the binary at the beginning of the mass transfer is large,
it begins when the companion evolves off the main sequence, the main
driving mechanism for mass transfer is nuclear evolution and
the system will evolve towards large orbital periods. Such systems are
said to be above the bifurcation period
 (Tutukov et
al. 1985, Ergma et al 1996). If the orbital period of the system at
the onset of the mass transfer is
small (e.g. it is below the bifurcation period), the companion 
is relatively unevolved, and 
the only important mechanism driving mass transfer are systemic angular 
momentum losses (AML) due to magnetic braking and gravitational radiation. This 
type of evolution is similar to the classical evolution of cataclysmic 
binaries. The orbital period becomes shorter and shorter and the systems  may 
experience a period gap when magnetic braking stops being effective (when the 
secondary becomes fully convective), and ultimately reaches a minimum period 
just before hydrogen burning is extinguished (Paczynski \& Sienkiewicz 1981; 
Rappaport, Joss, \& Webbink 1982). Beyond the period minimum (which depends on 
the evolutionary stage of the initial model), the donors radius begins 
increasing, following the mass- radius relation for degenerate stars, and the 
orbital period will increase again, driven by gravitational radiation alone. 
Actually, the distribution of orbital periods of LMXBs, based on relatively few 
systems (Liu, van Paradijs \& van den Heuvel, 2001), does not show a period gap 
as clear as that of CVs, but a more general limitation in the number of systems 
below $\sim 4$hr. It may well be that the distribution of initial periods of 
LMXBs favours the evolution of more evolved donors (Nelson \& Rappaport 2003) and population synthesis results show 
that several LMXB systems may be descendants of intermediate mass secondary, 
after a phase of thermal timescale mass transfer (Podsiadlowski,
Rappaport \& 
Pfhal 2002), but the evolutionary path towards shorter periods is mainly 
followed by evolutions starting from unevolved secondaries. For these systems, 
the donors have the same structure as the donors in CVs, and in principle there 
is no reason why they should not pass through a similar period gap, possibly 
having a smaller width, due to the fact that the accreting component is more 
massive (1.35 M$_\odot$) than the typical white dwarf in CVs ($\sim
0.6$ M$_\odot$) (Podsiadlowski, Rappaport \& Pfhal 2002).

Studies of the evolution of LMXBs usually disregard the compact nature
of the primary. This is not legitimate, since 
once the matter is transferred from the companion to the NS it
releases mass-energy due to the strong gravitational binding of the
NS, and therefore its 
gravitational mass (i.e. the ``charge'' of the gravitational force) decreases once it
is accreted onto the NS.
Thus even if the mass transfer is conservative
\textit{the total gravitational mass of a LMXB decreases} when matter is
transferred from the companion onto the NS:
it has been suggested that this could have observable effects
\cite{almo04}. Moreover, if enough mass is accreted
onto the primary, it can collapse to a black hole (Lavagetto et
al. 2004, hereafter Paper I). We should then keep into account both these effects if we want to study
accurately the evolution of a LMXB.

In this paper we will first of all introduce the evolution equations
for the binary system including the effects of general relativity
(section 2). Then we will show how  the mass-energy loss is potentially
measurable in binary millisecond radio pulsars (MSPs), and how this measurement can put
constraints on the equation of state (EOS) of NSs (section
3). Moreover we will show in section 4 under which conditions the NS
in systems below the
bifurcation period can survive the ``period gap''. Finally in section
5 we will study the evolution of a system above the bifurcation
period, showing how mass-energy loss alters the evolution
of the system.

\section{The evolution equations}

 We implemented a simulation code that includes
the stellar evolution of the companion, the evolution of the binary system
and the evolution of the NS under the effect of accretion
at the same  time in order to
simulate accurately the evolution of a LMXB. 
Our evolution code couples the routines of 
the ATON code (D'Antona, Mazzitelli and Ritter, 1988), updated with the
physical inputs described in (Ventura et al. 1998),
 which accounts for the stellar evolution and the binary evolution of
 the system, and routines accounting for the evolution of the NS (which 
is considered to be fully relativistic) under accretion\footnote{For a detailed description of
the fully relativistic study of the response of the NS to
the accretion of matter see Paper I.}.

If the primary of the system is a NS, its gravitational
mass will  be given by the sum of the baryonic mass (i.e. the number of 
baryons N times the average bare mass of the baryons $m_\mathrm{B}$)
and  of the potential and kinetic energies divided by $c^2$, that are non negligible since the gravitational
binding energy is large for matter as dense as neutron star matter. We know that for any
given equilibrium configuration of a NS we have on a general basis \cite{ba70}
\begin{equation}
  \label{eq:5mg}
  M_\mathrm{G} = M_\mathrm{G} (M_\mathrm{B}, J),
\end{equation}
where we indicate the gravitational mass of the star with $M_\mathrm{G}$, its
baryonic mass with $M_\mathrm{B}$ and its intrinsic angular momentum
with $J$. The accreted gravitational mass per unit time depends then both on
the number of baryons accreted and on the accreted angular momentum \cite{ba70}:
\begin{equation}
  \label{eq:5dmg}
  \dot{M}_\mathrm{G}= \left(\pder{M_\mathrm{G}}{M_\mathrm{B}}\right)_J \dot{M_B} + \left(\pder{M_\mathrm{G}}{J}\right)_{M_\mathrm{B}} \dot{J}.
\end{equation}
where
\begin{eqnarray}
  \label{eq:5bardeen}
  \left(\pder{M_\mathrm{G}}{J}\right)_{M_\mathrm{B}} &=& \f{\omega_\mathrm{NS}}{c^2}\\
\nonumber 
\left(\pder{M_\mathrm{G}}{M_\mathrm{B}}\right)_J &=& \Phi
\end{eqnarray}
where $\omega_{NS}$ is the NS spin frequency, $c$ is the
speed of light and $\Phi$ is the energy needed to bring a unit mass from infinity to
the pole of the star.

Therefore equation (\ref{eq:5dmg}) becomes:
\begin{equation}
  \label{eq:5dmgdt}
  \dot{M}_\mathrm{G}=  \Phi \dot{M_B} + \f{\omega_\mathrm{NS}}{c^2}\dot{J}.
\end{equation}
 
When matter is transferred from the companion onto the NS, and the
mass transfer is conservative, we have
\begin{equation}
  \label{eq:5mazz}
  \dot{M}_\mathrm{B}= - \dot{M}_\mathrm{c}.
\end{equation}
where $M_\mathrm{c}$ is the mass of the companion.
 It is useful to rewrite equation
(\ref{eq:5dmgdt}) using equation (\ref{eq:5mazz}) as \cite{almo04}
\begin{equation}
  \label{eq:5dmgdt2}
  \dot{M}_\mathrm{G}= -(1-\beta) \dot{M}_\mathrm{c}
\end{equation}
where $$0< \beta = 1 - \Phi - \f{\omega_{NS}}{c^2} \f{\dot{J}}{\dot{M}_\mathrm{B}}<1.$$
According to General Relativity the binding energy of the accreting
matter results therefore in a mass defect. The gravitational mass lost
in accretion is released from the system (mainly as X-rays) and
carries away a specific orbital angular momentum
that we assume to be equal to the specific orbital angular momentum of
the NS:
\begin{equation}
  \label{eq:lor}
  \left( \f{\dot{L}}{L}\right)_{\beta}= \frac{1}{L}\beta \dot{M}_\mathrm{c}
  \frac{2 \pi}{P} \left(a\frac{M_\mathrm{c}}{M_\mathrm{tot}}\right)^2
\end{equation}
where $P$ is the orbital period of the binary system and $a$ is the
orbital separation.
 Using the Kepler's law $$\frac{2 \pi}{P}= \left(\frac{G
     M_\mathrm{tot}}{a^3}\right)^{1/2}$$
where $ M_\mathrm{tot}= M_\mathrm{c}+ M_\mathrm{G}$, and the
expression for the orbital angular momentum
\begin{equation}
  \label{eq:orbi}
  L= M_\mathrm{c}M_\mathrm{G}\left(\frac{Ga}{
     M_\mathrm{tot}}\right)^{1/2} 
\end{equation}
we can rewrite
equation (\ref{eq:lor}) as
\begin{equation}
  \label{eq:lorgr}
   \left( \f{\dot{L}}{L}\right)_{\beta}= \beta
   \dot{M}_\mathrm{c}\frac{M_\mathrm{c}}{M_\mathrm{G} M_\mathrm{tot}}=
   \beta \frac{\dot{M}_\mathrm{c}}{M_\mathrm{c}
} \frac{q^2}{1+q}
\end{equation}
where $q=M_\mathrm{c}/M_\mathrm{G}$. The total variation of the orbital angualr momentum $L$ will then be equal to the sum of the systemic orbital angular momentum losses $
  \dot{L}_\mathrm{sys}$ and of the angular momentum losses due to the relativistic mass defect $\dot{L}_{\beta}$. We can therefore write:
\begin{eqnarray}
  \label{eq:5lorb}
\nonumber
 \left( \f{\dot{L}}{L}\right)_\mathrm{sys}& +&\left( \f{\dot{L}}{L}\right)_\mathrm{\beta} =\f{1}{2}\f{\dot{a}}{a}+
 \f{\dot{M_\mathrm{G}}}{M_\mathrm{G}} +
 \f{\dot{M_\mathrm{c}}}{M_\mathrm{c}} - \f{1}{2}\f{\dot{M_\mathrm{G}}
 + \dot{M_\mathrm{c}}}{M_\mathrm{G}+M_\mathrm{c}}\\
&=& \f{1}{2}\f{\dot{a}}{a}+ (1-q)\f{\dot{M_\mathrm{c}}}{M_\mathrm{c}}+
\beta \left[q - \frac{1}{2}\frac{q}{1+q}\right] \f{\dot{M}_\mathrm{c}}{M_\mathrm{c}}.
\end{eqnarray}
where we made use of equation (\ref{eq:5dmgdt2}).
If we now substitute equation (\ref{eq:lorgr}) into equation
(\ref{eq:5lorb}), we can write the derivative of the orbital
separation as:
\begin{equation}
  \label{eq:5a}
\nonumber  \f{\dot{a}}{a} = 2  \left( \f{\dot{L}}{L}\right)_\mathrm{sys} - 2
(1-q)\f{\dot{M_\mathrm{c}}}{M_\mathrm{c}}
 - \beta \frac{q}{1+q} \f{\dot{M_\mathrm{c}}}{M_\mathrm{c}} \\
\end{equation}
 The last term on the right is due to the relativistic mass defect: it
 relevant only in compact systems where $\beta$ is
 non-negligible. Using Kepler's Law, we can write for the evolution of
 the orbital period:
\begin{eqnarray}
  \label{eq:5p}
\nonumber  \f{\dot{P}}{P} &=& \f{3}{2} \f{\dot{a}}{a} - \f{1}{2}\f{\dot{M_\mathrm{G}}
 + \dot{M_\mathrm{c}}}{M_\mathrm{G}+M_\mathrm{c}} \\
&=& 3  \left( \f{\dot{L}}{L}\right)_\mathrm{sys} - 3
(1-q)\f{\dot{M_\mathrm{c}}}{M_\mathrm{c}}- 2 \beta \frac{q}{1+q} \f{\dot{M_\mathrm{c}}}{M_\mathrm{c}}
\end{eqnarray}
where the third term appears due to the relativistic mass deficit. This term is
positive since $-\dot{M}_\mathrm{c}$ is positive: this means that as mass is transferred, an addictional positive contribution to the orbital period derivative is present in relativistic systems.
 One should keep in mind that the mass transfer rate $\dot{M}_\mathrm{c}$ depends upon
the whole evolution of the binary system: in general it is a function of
the nuclear evolution of the companion and  of the orbital
separation. In order to evaluate quantitatively the 
influence of the evolution of the NS on the evolution of a system we
must
carry out detailed numerical simulations of the binary system due
to the non-linearity of the equations.
It is straightforward to extend these equations to the case of non-conservative mass transfer.

In the next sections, we will show how these effects, together with
the evolution of the compact object, can both have observable effects
and alter the secular evolution of LMXBs.

\section{Observable effects of general relativity}

We may ask ourselves if the effects of relativity on the orbital
parameters of the binary system 
can be observable. In theory, any binary system about which we have
enough information can reveal the effects we described in the
preceding section. Anyway, since we do not know much about most binary
systems and, as various effects can overlap (see for example section 5), we
have little chance to observe directly these effects, which they only
play a role in the  secular evolution of the
system.
Al\'ecian \& Morsink (2004) argued that if the orbital period
derivative of an accreting LMXB can be measured, it will allow to
derive information on the structure of the NS, such as its mass and
its gravitational binding energy. 
This effect can anyway be measurable only if the mass transfer is
driven exclusively by the emission of gravitational
radiation: there are too many uncertainties on the amount of
angular momentum lost due to magnetic braking and on the nuclear
evolution of the companion to allow to separate the relativistic
effects in the orbital evolution in such cases. This
limits the range of systems that can be interesting to close LMXBs
(with $P \la 2$ h) with a main sequence companion. In this situation the mass loss
from  the system will result in a potentially observable modification of the orbital period
derivative. As Al\'ecian \& Morsink point out, however, uncertainties in the
physics of binary evolution and of mass accretion make it
difficult to separate this effect on the orbital period from others.
Moreover, it is impossible to infer
 the mass accretion rate in a LMXB from its observed X--ray
luminosity with the accuracy that is needed  to extract information in
their model. It is therefore unlikely that this
effect can be  used to investigate the EOS of ultradense matter.

On the other hand, when accretion onto the primary ends, the NS lights up as a radio
pulsar, and  it brakes down due to rotating magnetodipole emission.
The NS loses some of its gravitational mass because it
is radiating away energy. We can then write equation (\ref{eq:5p}) in the form
\begin{equation}
  \label{eq:5pulsar_porb}
  \f{\dot{P}}{P}= -2 \frac{\dot{M}_\mathrm{G}}{M_\mathrm{c}}  \frac{q}{1+q}
  +\f{\dot{P}_{GW}}{P} 
\end{equation}
where $\dot{P}_{GW}/P =3 \dot{L}_{GW}/L$ is the orbital period derivative due to the
emission of gravitational waves from the binary system, that we can
write as \cite{lo01}:
 \begin{eqnarray}
   \label{eq:5lorimer}
   \dot{P}_\mathrm{GW} &=& - \f{192 \pi}{5 c^5} \left( \f{2 \pi G}{P}
   \right)^{5/3} \f{M_\mathrm{G} M_\mathrm{c}}{M_\mathrm{tot}^{1/3}}\\ \nonumber &\cdot &\left(1 + \f{73}{24} e^2 + \f{37}{96}
   e^4\right) (1-e)^{-7/2}
 \end{eqnarray}
where  $e$ is the eccentricity of the system.
 It is then
evident that when enegry is released from the MSP,
the mass defect yelds a positive contribution to the orbital period derivative,
 opposite to the contribution of gravitational waves emission. Since the effect of
gravitational mass loss is $\propto P$, while the effect of
gravitational waves emission is $\propto P^{-5/3}$, the former effect
will be dominant in systems with large enough orbital periods (say $P \ge 6$
h), while the  latter will become more relevant in systems with short periods.

 The variation of the
gravitational mass is equal to the spin-down energy of the pulsar,
divided by $c^2$. To a good approximation we can write:
\begin{equation}
  \label{eq:enloss}
  \dot{M}_\mathrm{G} = \frac{I}{c^2} \omega_\mathrm{NS} \dot{\omega}_\mathrm{NS}
\end{equation}
where I is the moment of inertia of the NS.
Esposito and  Harrison (1975)
readily noticed after the discovery of the first binary pulsar,
PSR~1913+16, that the mass defect can alter the orbit of a binary
pulsar. They found that its
effect on the orbital evolution of PSR~1913+16 was
negligible. Today we know many MSP, that have periods below 5
ms and whose spin-down mass loss is larger (see equation \ref{eq:enloss}).
In these systems, the effect can be orders of magnitude stronger that
the effect of gravitational waves. When this is the case, measuring
the orbital period derivative of the binary system can help to put
strong constraints on the EOS of NSs. 
 In many cases it is possible to measure both the
 spin frequency and its derivative in a radio pulsar with high
 precision. Moreover,  the absence of
 mass transfer cuts away any uncertainty in the binary evolution model. 
The orbital period derivative depends then only
on measured quantities ($\omega_{NS}$,$\dot{\omega}_{NS}$ and $e$), on the
masses of the two stars and on the moment of inertia of the neutron
star (see equations \ref{eq:5pulsar_porb}, \ref{eq:5lorimer} and \ref{eq:enloss}). We can get
information on the two masses from the mass function
$$ f(M) = \f{M^3_\mathrm{c} \sin^3 i}{M_\mathrm{tot}^2}$$
 that is measurable in binary radio pulsars with very good precision.
 Using it, we can impose constraints on the moment of inertia of
 the NS. Since the moment of inertia depends strongly on the EOS of
 the NS \cite{cost94}, the detection of this effects will allow us to
discriminate between various EOSs on a solid observational basis.

\begin{figure}
  \centering
  \epsfig{figure=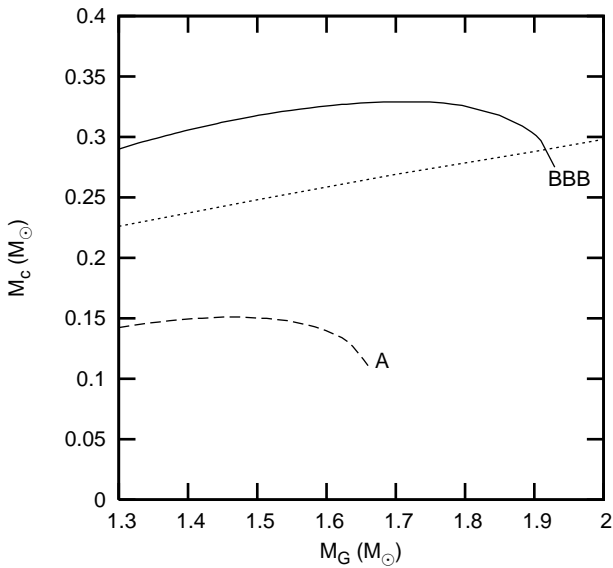}
  \caption{Allowed values of the mass of the primary versus the mass
    of the companion (both in solar masses) for a system orbital
    period  $P= 8$ h, a spin period $P_s = 2$ ms, $\dot{P}_s = 3
    \times 10^{-19}$  and an orbital period derivative of $2.5 \times
    10^{-14}$. The two lines in the figure are for NSs with
    EOS A  (dashed line) and NSs with EOS BBB (solid line). The dotted line indicates
    the lower limit on the companion mass obtained if the mass
    function is $5 \times 10^{-3} M_\odot$.}
  \label{fig:5pulzar}
\end{figure}

Let us show this method with an example:
 suppose that we will observe a system with an orbital
period $P= 8$ h, a spin period $P_s = 2$ ms, $\dot{P}_s = 3 \times
10^{-19}$, a mass function of $5 \times 10^{-3} M_\odot$ and that we
have measured  the orbital
period derivative to be $+2.5 \times 10^{-14}$. In figure
\ref{fig:5pulzar} we plot the values  of the masses
of the two stars  that are compatible with this value of the orbital
period derivative, in the hypothesis that the NS is governed by
 the pure neutron EOS by Pandharipande, named EOS A in the classic
 catalog by Arnett \& Bowers (1977), or  by the realistic hadronic EOS
 by Baldo, Bombaci and Burgio (1997), that we label as EOS BBB.
The mass function imposes that the values of the two masses
should be above the dotted line in figure \ref{fig:5pulzar}. The  figure tells us that,
given the mass function and the orbital period derivative we assumed,
the pulsar cannot be described by EOS A.
 
Now that we have shown how promising is this effect in principle, we
may ask ourselves which is, between the known MSPs, 
the best candidate for detecting such an effect? 
The most promising object we found is
PSR~J0218+4232, which is constituted of a NS spinning at $2.3$ ms  in orbit with a white
dwarf companion, with an orbital period of 2
days . The mass ratio is measured to be $q=0.13 \pm 0.04$ \cite{bavk03}.

Due to its strong power output, the pulsar loses gravitational mass at the rate $4 \times 10^{-12} I_{45}$
 M$_\odot$/ yr,  where $I_{45}$ is the moment of inertia of the NS in
 units of $10^{45}$ g cm$^{2}$. The corresponding orbital period
 derivative, according to equation (\ref{eq:5pulsar_porb}), becomes
 $\dot{P}_{orb}=2.5 \times 10^{-14} I_{45}$. The variation
 in the orbital period is derived from the measure of
the periastron time delay. The effect of an orbital period derivative
on the periastron arrival time is given by:
 \begin{equation}
   \label{eq:5dtperi}
   \Delta T_\mathrm{per} \simeq 0.5 \f{\dot{P}}{P} \Delta T^2_\mathrm{obs}.
 \end{equation}
We find that we will need $117/I^{1/2}_{45}$ years of observation to
detect a delay of 1 second, thus rendering impractical to measure such
an effect in this pulsar, although period derivatives of the same
order of magnitude have already been detected (see e.g. Nice et al. 2004),
but in pulsars with short orbital periods. Since the relativistic
effect becomes dominant in pulsars with a large enough orbital period,
we should seek for a pulsar with an higher spin-down energy -- and
therefore with an higher orbital period derivative for the same
orbital period -- in order to
measure this effect in a shorter observation time. Such a
measurement is then quite unrealistic for presently known millisecond
pulsars. However if we will find a pulsar with sufficiently high spin-down
power it can become observable in a decade, becoming therefore feasible.

\section{How are short period LMXBs formed?}

\begin{figure}[h]
  \centering
\epsfig{figure=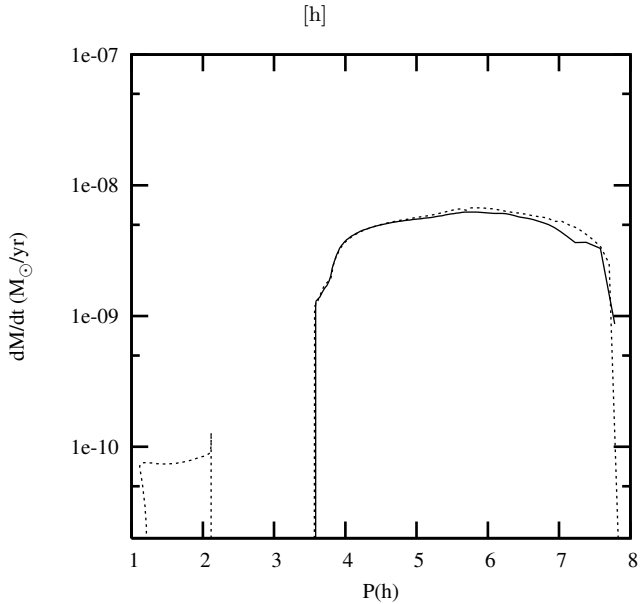}
  \caption{Mass accretion rate as a function of the orbital period for
    a system below the bifurcation period. The companion is a
    population I star  of $1 M_\odot$. We show the evolution of the system,
    not taking into account the evolution of the NS (dotted line) and
    the evolution of the system when we include the evolution of the primary:
    the companion transfers most of its mass due to magnetic braking,
    passing from $P\sim 8$ h to $P \sim 3.5$ h when the star becomes
    fully convective and mass transfer stops. The mass transfer then
    resumes once the system is close enough, at $P\sim 2$ h in the
    system with an unevolved primary. When we keep into acocunt the
    evolution of the primary, the system evolves similarly to the
    classical one, but when the companion becomes fully convective the NS is
    supramassive and therefore it collapses to a black hole before
    accretion can resume. As can be seen from the figure, no real
    difference is present in the evolution above the period gap, even
    if the relativistic mass defect is kept into account. For the sake
    of claryty, data were smoothed clean the numerical noise.}
  \label{fig:restrrel}
\end{figure}

Let us consider a binary system that is below the bifurcation period,
 and is constituted of a companion of $1 M_\odot$  and of a primary with an initial gravitational mass
of $1.35 M_\odot$  -- a mass that appears to be
the typical of isolated NSs \cite{toch99}. The orbital period of the binary at the onset of
the mass transfer is $\sim 8$ h.
 In the standard scenario, the system transfers mass because it loses
 angular  momentum due to magnetic braking of the companion, at a rate
 large enough to push the star well out of thermal equilibrium for
 periods shorter than $\sim 4$ h. 
When the companion becomes fully
convective (typically at an orbital period $\sim$ 3
h), magnetic braking is thought to stop, the companion recovers
thermal equilibrium and the mass transfer ceases
\cite{spri83}. The binary system will continue to shrink due
to gravitational waves emission, until it reaches an orbital period
$\sim$ 2 h, when mass transfer resumes. The evolution of this system
in the period vs. accretion rate plane is shown in panel a of figure
 \ref{fig:restrrel}.  We used the magnetic braking law by Verbunt \&
 Zwaan (1981) with a braking index of 0.5. In doing this first
 simulation,  we neglected any evolution of the NS. 

What happens to the system if we keep into account the evolution of
the primary?
The fate of the NS in the period gap happens to be very
interesting. An average $\sim 0.8 M_\odot$ have been accreted, and the
NS is therefore rapidly spinning (see Paper I), and it will light up as a
millisecond radio pulsar since accretion has stopped. Two general
relativistic effects are important in this phase:
\begin{enumerate}
\item The loss of gravitational mass
  from the pulsar yields an additional positive contribution to the
  orbital period derivative (see the preceding section). This
  additional effect contrasts the shrinking of the orbit due to the
  gravitational waves emission, thus
  increasing the duration of the detached phase of the
  system. This increase can vary strongly depending on the spin-down
  energy of the primary and on $q$ (see equation
  \ref{eq:5pulsar_porb}). 
\item The silent collapse to a black hole if the pulsar is supramassive
  (i.e. its mass exceeds the maximum non-rotating mass), once it loses
  enough angular momentum.
\end{enumerate}
For the NS to survive the gap, the timescale of the collapse ($T_c$),
must be larger than the timescale needed for the system to ``cross''
the period gap ($T_g$). Else, the NS will collapse to a black hole
before the mass transfer resumes.
This happens, for example, in our simulated system if the  primary is
a NS governed by EOS BBB,
as can be seen in panel b of figure \ref{fig:restrrel}.

The collapse time $T_c$ is defined by the equation
\begin{equation}
  \label{eq:timedip}
  J_\mathrm{in} - J_\mathrm{crit} = - \int_0^{T_c} \dot{J} dt
\end{equation}
where $J_\mathrm{in}$ is the angular momentum at the beginning of the
detached phase, $J_\mathrm{crit}$ is the critical angular momentum below which
the star collapses, and $\dot{J}$ is the angular momentum lost during
the pulsar phase.  Combining equation
(\ref{eq:5dmgdt}) and the formula for the energy released by a
magnetodipole rotator in general relativity \cite{ream04}, $\dot{J}$ is given by
\begin{equation}
  \label{eq:5evpul}
  \dot{J} = - \f{2}{3 c^3} \mu^2 \omega_{NS}^3 \left(\f{f}{N^2}\right)^2
\end{equation}
where
\begin{eqnarray}
  \label{eq:5rezzolla}
  N&=&\left(1 - 2\chi \right)^{1/2}, \hskip
  0.5cm \chi=\f{G M_\mathrm{G}}{c^2 R_{NS}}\nonumber \\
  f&=& \f{3}{8} \chi^{-3} \left[\log{N^2} + 2\chi \left( 1+
  2\chi \right) \right].
\end{eqnarray}

The gap time $T_g$ is defined by the equation 
\begin{equation}
  \label{eq:timeg}
  \Delta P_\mathrm{gap} = - \int_0^{T_g} \dot{P} dt
\end{equation}
where $\Delta P_\mathrm{gap}$ is the amplitude of the period gap and
$\dot{P}$ is defined in equation (\ref{eq:5pulsar_porb}). In
most situations $\Delta P_\mathrm{gap} \sim 1$ h,
and the mass of the companion
is $\la 0.25 M_\odot$. Integration of equations (\ref{eq:timeg}) and 
(\ref{eq:timedip}) shows that $T_g > T_c$ if the NS is
supramassive and if the dipole magnetic field of the neutron
star exceeds $10^{26}$ G. This result holds for a vast range of EOSs
from softer ones like EOS A  to the stiffer ones like EOS L
\cite{ab77}, including recent, realistic EOSs like EOS FPS
\cite{lora93}, EOS APR \cite{apr} and EOS BBB \cite{babobu97}, 
as long as we assume  that the mass of the NS exceeds of at
least $0.02 M_\odot$ the maximum non-rotating mass. The result holds independently
from the angular momentum of the NS at the onset of the pulsar phase. 
This means that LMXBs that host a NS and have a period $\la
2$ h, like all the millisecond X-ray pulsars known to date, cannot be
supramassive if they have a non-negligible magnetic field. This result
obviously holds only if the system is a NS-MS (main sequence) binary
that has evolved through the period gap.

 For instance, the surface magnetic field of the first millisecond
 X-ray pulsar discovered, SAX~J1808.4-3658 \cite{wiva98}, has been estimated to be in the
range $(1-5) \times 10^8$ G \cite{dibu03}.
If this system has a MS companion, so that it evolved from longer
orbital periods, it cannot be supramassive, as it
survived the period gap.  The primary will not be a supramassive NS
only  if one of the following holds:
\begin{enumerate}
\item A relevant
part of the accreting matter has been ejected from the system, and the
mass transfer has therefore been non-conservative for most of the
binary evolution.
\item The maximum non-rotating mass of the NS is very high, $\ge 2 M_\odot$
(i.e. the EOS of the NS is very stiff). 
\end{enumerate}

\section{Secular evolution of systems above the bifurcation period}

In the preceding section, we have shown how the evolution of the
compact object can alter the evolution of a system below the
bifurcation period. Now we will show how the evolution of the NS can
alter the secular evolution of a system above the period gap.
In such a system, the mass transfer is driven by the nuclear evolution
of the companion.The evolution of such wide systems
is well understood  \cite{wrs83}, and is thought to explain well the
formation of binary MSPs with large orbital periods

In order to show how this canonical evolution is altered when we keep
into account the evolution of the primary, we simulated a binary
system whose companion is
a $1.1 M_\odot$ population I star. The initial mass of the primary is
chosen again to be  1.35 M$_\odot$.
When the mass transfer begins the orbital period of the system is 11 d.  

First of all, we carried out a simulation in which the we disregarded
the evolution of the primary, for which we assumed $M_\mathrm{G} = M_\mathrm{B}$.
This system is indicated in the following as system A.
 We simulated also a system
(named B in the following) with the same companion star, but
this time taking into account the evolution of the primary. The NS was
assumed to have a low surface magnetic field of $10^8$ G, and
the EOS has been fixed to be EOS BBB we already introduced in section 3.
 In paper I we showed that a weakly magnetized NS is easily spun up to
periods well below one millisecond, uncomfortably lower than the
minimum observed period for a NS, 1.56 ms \cite{baal82}.
There is mounting evidence that some mechanism, whose nature is still
unclear, prevents NSs to spin faster than 700 Hz \cite{chal03}. We
simulated therefore a system identical to
system B, but we kept the spin frequency artificially below 700 Hz. We
will refer to it as system C.

As we said before, the evolution of the mass of a NS differs from
that of a non-collapsed star because of the discrepancy between
the gravitational and the baryonic masses in NSs. According to
equation  (\ref{eq:5dmgdt})  the gravitational
mass of the NS will be smaller if the spin frequency and
the angular
momentum of the star are smaller for a given baryonic mass. 
Therefore, we expect that the mass of the NS in system C will be
smaller than the one of the NS in system B, and both will smaller
than that of system A.
From our simulations we find in fact that 
the gravitational mass of the NS at the end of the
evolution is $1.96$ M$_\odot$ in system A, $1.88$ M$_\odot$ ($4 \%$
less)  in system
B, and  $1.79$ M$_\odot$ ($9\%$ less) in system C. The higher mass
defect in system C is due both to the smaller angular momentum of the
star  (that is almost
irrelevant, since in a neutron star the rotational energy is usually
$\sim 0.1$ times the binding energy) and to the fact that when the NS spins slower it is more compact
and therefore its binding energy is also higher.

\begin{figure}
  \centering
  \epsfig{figure=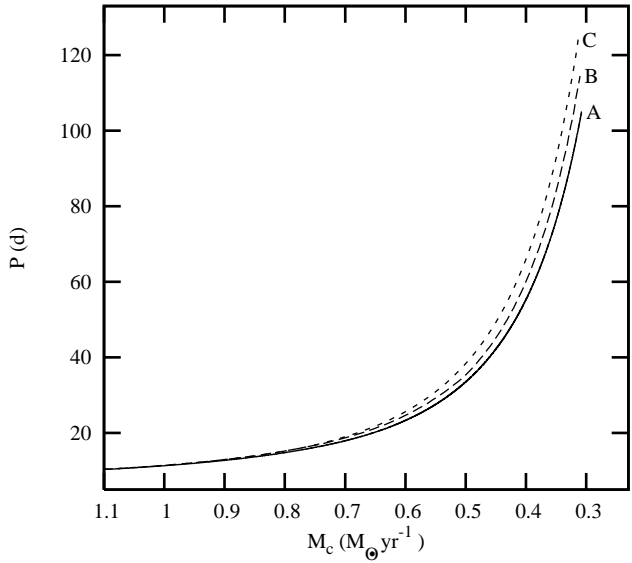}
  \caption{Orbital period (in
    days) of the binary  system as a
  function of the mass of the companion (in solar masses).
 Letters indicate the evolutionary
  tracks for systems A,B and C. The evolution is followed only until the
  companion collapses to a white dwarf. The x axis is inverted, so
  that the evolution of the systems goes from left to right.}
  \label{fig:period}
\end{figure}

It is interesting to note that the evolution of the NS has other
effects on  the evolution of the binary system.
 We know that the final state of such systems
depends substantially only on the mass of the inner core of the
companion star \cite{wrs83}.
Looking at equation (\ref{eq:5p}) we see that the relativistic effect
is dominant during the first phases of accretion, when $q \sim
1$: when matter is transferred, a relativistic
system enlarges faster than a non-relativistic one. Consequently, the Roche Lobe of the companion is
larger, and the mass accretion rate will become smaller. As a consequence, the
relativistic systems have a longer time to evolve, and the mass of the
core becomes slightly bigger: at the end of accretion, the companion
in system A has a core of $0.305 M_\odot$. In system B the
companion has a core that is $0.003 M_\odot$ heavier than in system
A, while in system C, the core is $0.006 M_\odot$ heavier. As a
consequence (see again Webbink et al. 1983), the
orbital period at the end of the accretion process 
is 105 d in system A, $114.5$ d ($\sim 9 \%$ larger) in system B and
124 d ($\sim 18 \%$ larger) in system C (as can be seen in figure \ref{fig:period}).  

This means  that the net effect on the orbital period evolution of the
system is stronger than the effect on the mass alone, but this is due
to the small changes in the final core mass and not directly to the
third term on the right of equation (\ref{eq:5p}). Small changes in the mass
of the core also account for significant variations in the orbital period
at the end of mass transfer: this means that, although the effects we
describe can change the evolution of a system from given initial conditions,
it does not alter the scenario of the evolution of
such systems.

\section{Conclusions}

In this paper we showed that the presence of a NS, that can be
described using  general relativity, can have a big
impact on the evolution of the binary system.

First of all we searched for potentially observable effects of the
relativistic nature of the primary. We noticed that when the NS releases energy
without accreting during the pulsar phase, it will lose gravitational
mass. Therefore, a positive contribution to the orbital period
derivative can dominate in certain situations over the negative
contribution due to gravitational waves emission, resulting in an
overall \textit{positive} orbital period derivative.  
We showed that the measurement of such a period
derivative in a binary millisecond pulsar can allow to put
 constraints on the EOS of ultradense matter on a solid
observational basis.

Then we concentrated on the effects that the secular evolution of the
primary can have on the evolution of binary systems, both below and
above the bifurcation period. 
In systems starting below the bifurcation period a
necessary condition for a NS to survive the period gap without
collapsing to a black hole is to be
non-supramassive. This means
that the NS in LMXBs with short periods, like the millisecond X-ray
pulsar SAX~J1808.4-3658, cannot be supramassive if they evolved from
larger periods. This implies that either the mass transfer is highly non
conservative or the EOS of ultradense matter is stiff.

In systems starting above the bifurcation period, we showed how the relatively small
effect of General Relativity on the orbital evolution can alter the
evolution of the companion; this means that the total effect can be
non-negligible. However these effects do not change the proposed
scenario for the evolution of large-period systems in any significant way.

In all, the relativistic mass defect has some effect on the evolution
of binary systems, but it is negligible if compared with the
uncertainties that we still have on the other effects present in the
standard theory of the binary evolution. For example, various laws
have been proposed for the magnetic braking mechanism, and the
intensity of the effect has been questioned (see e.g. Ivanova
\& Taam, 2003). However, the evolution of the primary - and the
resulting mass defect - can be important in studying the evolution of
the binary system, both because it can give rise to an observable
effect in MSPs and because it puts constraints on the evolution of
systems below the bifurcation period, and therefore cannot be
disregarded when doing evolutionary studies.

\end{document}